# Voltage Controlled Magnetic Skyrmion Motion for Racetrack Memory

Wang Kang[1,2,3], Yangqi Huang[1,2], Chentian Zheng[1], Weifeng Lv[3], Na Lei[1], Youguang Zhang[1,2], Xichao Zhang[4], Yan Zhou[4,6†], and Weisheng Zhao[1,2,5,*]

[1] *Fert Beijing Institute, Beihang University, Beijing, China*
[2] *School of Electronic and Information Engineering, Beihang University, Beijing, China*
[3] *School of Computer Science and Engineering, Beihang University, Beijing, China*
[4] *Department of Physics, University of Hong Kong, Hong Kong, China*
[5] *Institut d'Electronique Fondamentale (IEF), Univ. Paris-Sud, CNRS, Orsay, France*
[6] *School of Electronic Science and Engineering, Nanjing University, Nanjing, China*

*E-mail: weisheng.zhao@buaa.edu.cn

†E-mail: yanzhou@hku.hk

**Abstract**

Magnetic skyrmion, vortex-like swirling topologically stable spin configurations, is appealing as information carrier for future nanoelectronics, owing to the stability, small size and extremely low driving current density. One of the most promising applications of skyrmion is to build racetrack memory (RM). Compared to domain wall-based RM (DW-RM), skyrmion-based RM (Sky-RM) possesses quite a few benefits in terms of energy, density and speed *etc*. Until now, the fundamental behaviors, including nucleation/annihilation, motion and detection of skyrmion have been intensively investigated. However, one indispensable function, *i.e.*, pinning/depinning of skyrmion still remains an open question and has to be addressed before applying skyrmion for RM. Furthermore, Current research mainly focuses on physical investigations, whereas the electrical design and evaluation are still lacking. In this work, we aim to promote the development of Sky-RM from fundamental physics to realistic electronics. First, we investigate the pinning/depinning characteristics of skyrmion in a nanotrack with the voltage-controlled magnetic anisotropy (VCMA) effect. Then, we propose a compact model and design framework of Sky-RM for electrical evaluation. This work completes the elementary memory functionality of Sky-RM and fills the technical gap between the physicists and electronic engineers, making a significant step forward for the development of Sky-RM.

## Introduction

Magnetic skyrmion, which has been discovered in bulk ferromagnets, magnetic thin films and nanowires, are topologically stable nanoscale magnetization configurations with particle-like properties [1-12]. The origin of skyrmion can be explained as a consequence of competitions between the ferromagnetic exchange coupling, magnetic anisotropy and Dzyaloshinskii-Moriya interaction (DMI, either bulk or interfacial) in magnetic systems lacking inversion symmetry [13-16]. Here the bulk (or interfacial) DMI between two atomic spins $S_1$ and $S_2$ with a neighboring atom in lattices (or heavy metal in magnetic thin films) having a large spin-orbit coupling (SOC) can be expressed as $H_{DM} = -D_{1,2} \cdot (S_1 \times S_2)$, where $D_{1,2}$ denotes the DMI vector. Experimental observations of skyrmions in the bulk materials (*e.g.,* MnSi and FeGe) have been reported early in 2009 [1]. Recently, the skyrmions induced by interfacial DMI in ferromagnetic thin films have also been observed on Fe monolayer (or PbFe bilayer) on Ir [6]. Owing to the promising merits of skyrmion, such as topological stability, small size and ultra-low spin-polarized current density needed to move them, tremendous research efforts have been made for exploring them in future spintronic applications [17-22]. One of the potential applications of skyrmion is to build racetrack memory (RM), in which binary data information is encoded in the form of skyrmions and transmitted in a nanowire [19-20]. In comparison with domain wall- based RM (DW-RM) [23, 24], where data information is encoded by a train of spin-up and spin-down magnetic domains separated by DWs, skyrmion-based RM (Sky-RM) is promising to achieve lower energy cost, higher device packing density and more robust data stability.

Until now, the fundamental behaviors, including nucleation/annihilation, motion and detection of skyrmion have been intensively studied. Nevertheless, to employ skyrmion in real RM application, the manipulation of skyrmion, *e.g.*, pinning/depinning at a specific position of the nanotrack is indispensable. To our best knowledge, this is still an open question. In addition, to implement and eventually commercialize Sky-RM, an electrical model and design methodology are also indispensable. In this work, we aim to promote the development of Sky-RM from fundamental physics to realistic electronics and to bridge the physicists and electronic engineers. First, the pinning/depinning characteristics of Sky-RM through voltage-controlled magnetic anisotropy (VCMA) were studied in detail by micromagnetic simulations. Then, a physics-based electrical model of Sky-RM was developed based on current experimental results and theoretical explanations. Finally, the evaluation and optimization of Sky-RM were investigated from both physical and electronic perspectives. This study makes a significant step forward for the development of Sky-RM and will trigger more theoretical, experimental and engineering investigations in the promising research direction of *Skyrmionics*.

## Results

**Sky-RM structure**. Figure 1 shows the schematic diagram of the Sky-RM, which is composed of five parts: write head for skyrmion nucleation, nanotrack (racetrack) for skyrmion motion, read head for skyrmion detection, VCMA gate for skyrmion pinning/depinning as well as the peripheral complementary metal-oxide-semiconductor (CMOS) circuits for generating the nucleation current ($I_{nucl}$), driving current ($I_{driv}$) and detection current ($I_{det}$). The skyrmion is initially nucleated by injecting local spin-polarized current ($I_{nucl}$) through the spin-valve (*e.g.,* Co/Cu/CoFeB/Ta/Co/[Co/Pt]$_n$) write head (with diameter of 20 nm and the spin-polarization factor of $P = 0.4$). Then it moves along the nanotrack (*e.g.,* Pt/Co thin films, with length × width ×thickness of 1000 nm × 80 nm × 0.4 nm) by a vertical spin current ($I_{driv}$). In actual simulations, the length of the nanotrack varies for saving simulation time. During the motion, the skyrmion can be pinned/depinned at the VCMA gate (with width of 60 nm) by modulating the magnetic anisotropy. Finally, the skyrmion can be detected by applying a detection current ($I_{det}$) through the magnetic tunnel junction (MTJ) read head (with diameter of 40 nm) because of the tunnel magneto-resistance (TMR) effect. The nucleation, motion and detection of skyrmion have been intensively studied in literatures [10,17,19-20]. Here we focus mainly on the pinning/depinning characteristics of skyrmion by using the VCMA effect.

**Pinning/depinning of skyrmion using the VCMA effect.** As we know that the perpendicular magnetic anisotropy (PMA) can be locally modulated by applying a voltage because of the charge accumulations [22,25,26]. This effect can be employed for pinning/depinning of skyrmion in RM application. In this section, we investigate the operation conditions and characteristics of skyrmion pinning/depinning by VCMA as well as the design strategy for Sky-RM. In the micromagnetic simulations, we set the default parameter values of the DMI of $D = 3 \text{ mJ/m}^2$, PMA of $K_u = 0.8 \text{ MJ/m}^3$ and Gilbert damping of $\alpha = 0.3$, which are corresponding to the material parameters of Co/Pt films [17,27]. The VCMA effect is based on a linear relationship [25], *i.e.,* $K_{uv} = K_u + \vartheta V_b$, where $V_b$ (from $-2$ V to $+2$ V) is the bias voltage on the VCMA gate and $\vartheta$ (with default value of 0.02) is a coefficient.

Figure 2a shows the schematic view (x-y plane) of the skymion states when passing the VCMA-gated region. At the VCMA-gated region, the change of PMA induces energy barrier on the boundaries at the left side and right side of the region. For simplicity, if the PMA of the left side of the boundary is larger than that of the right side, we denote a negative energy barrier ($-\Delta E_b$), otherwise, we denote a positive energy barrier ($+\Delta E_b$). Obviously, each VCMA-gated region has a pair of energy barriers, *i.e.,* $\{+\Delta E_b, -\Delta E_b\}$ or $\{-\Delta E_b, +\Delta E_b\}$ depending on the applied voltage polarity (see Figure 2b). Given a specific driving current and a VCMA gate, whether or not a skyrmion can pass the VCMA-gated region depends on the competition between the driving current and the positive energy barrier ($+\Delta E_b$). Figure 2c shows the working window for the baseline $K_u = 0.8 \text{ MJ/m}^3$ at various voltages (or energy barriers) and driving current densities. There are four cases: (a) obviously, if the driving current is off, the skyrmion cannot pass the VCMA-gated region; (b) the energy barrier is $\{+\Delta E_b, -\Delta E_b\}$, the driving current is on but its driving force cannot overcome $+\Delta E_b$, the skyrmion stops at the left boundary of the VCMA-gated region (see Supplementary Figure S1-a); (c) the energy barrier is $\{-\Delta E_b, +\Delta E_b\}$, the driving current is on but its driving force cannot overcome $+\Delta E_b$, the skyrmion stops at the right boundary of the VCMA-gated region (see Supplementary Figure S1-a); (d) the driving current is on and it is able to overcome $+\Delta E_b$, the skrymion passes the VCMA-gated region (see Supplementary Figure S1-b and S1-c). Figure 2d shows the working window for a constant energy barrier ($|K_{uv} - K_u| = |\Delta E_b| = 0.04 \text{ MJ/m}^3$) at various baseline $K_u$ and driving current densities. We find that the pinning/depinning of the skyrmion depends only on the potential height of the energy barrier. Therefore, a smaller $K_u$ is preferable for Sky-RM in practice to achieve high skyrmion motion velocity. In addition, it should be noted that if the driving current density is too large (e.g., $j_{driv} = 6 \text{ MA/cm}^2$), the skyrmion may reach the nanorack edge and be annihilated because of the transverse motion (see Supplementary Figure 2a-2b). In the following, we choose a baseline $K_u = 0.8 \text{ MJ/m}^3$ and two typical $K_{uv}$ (i.e., $1.05 K_u$ and $0.95 K_u$) for detailed evaluations.

In order to employ the VCMA effect for pinning/depinning of skyrmions in the Sky-RM, two strategies are investigated. The first one is to directly control the (on/off) voltage of the VCMA-gated region. In this case, the driving current (e.g., $j_{driv} = 2 \text{ MA/cm}^2$) is a consistent DC current and its driving force is not enough to overcome the energy barrier of the VCMA-gated region when the voltage is on. Figure 3 shows the trajectory of the skyrmion motion along the nanotrack. The skyrmion stops either at the left boundary (for $V_b = +2.0 \text{ V}$ and $K_{uv} = 1.05 K_u$) or at the right boundary of the VCMA-gated region (for $V_b = -2.0 \text{ V}$ and $K_{uv} = 0.95 K_u$) when the voltage of the VCMA gate is on and then continues to move when the voltage is off (see also Supplementary Movie 1 and Movie 2). By placing the VCMA gates evenly along the nanotrack, this strategy is rather suitable for Sky-RM applications with step-by-step motion based on the clock frequency. The second strategy is to set the $K_{uv}$ of the VCMA-gated region to be a constant. In this case, we modulate the driving current dynamically to manipulate the pinning/depinning of skyrmion. As shown in Figure 4, the total driving current is composed of a DC component and an AC pulsed current. The current density (e.g., $2 \text{ MA/cm}^2$) of the DC component is limited and its driving force cannot overcome the energy barrier of the VCMA-gated region. Therefore the skyrmion stops either at the left boundary (for $V_b = +2.0 \text{ V}$ and $K_{uv} = 1.05 K_u$) or at the right boundary of the VCMA-gated region (for $V_b = -2.0 \text{ V}$ and $K_{uv} = 0.95 K_u$) if the AC current pulse is off. However, when the AC current is on, the total amplitude of the driving current (e.g., $4 \text{ MA/cm}^2$) is large enough to overcome the energy barrier of the VCMA-gated region. In this case, the skyrmion can pass the VCMA-gated region if the pulse duration of the AC current is long enough (see Supplementary Movie 3 and Movie 4). This strategy offers more freedoms for Sky-RM design by dynamically configuring the driving current. As can be seen, when the skyrmion approaches or even half crosses the positive energy barrier ($+\Delta E_b$) of the VCMA gate under the driving current, it stops and moves back in the longitudinal direction due to the repulsive force of the energy barrier if the driving current cannot overcome the energy barrier. For each strategy, the voltage or AC current pulse width required for skyrmion to pass the VCMA-gated region varies. Specifically, the pulse width for the case of $K_{uv} = 1.05 K_u$ is much longer than that for the case of $K_{uv} = 0.95 K_u$, as the skyrmion motion velocity inversely proportional to the PMA value.

**Sky-RM implementation and optimization from an electronic perspective.** In order to investigate the electrical characteristics of the Sky-RM, we developed a physics-based electrical model based on current experimental results and theoretical explanations.

The skyrmion nucleation mechanism has been widely studied in literatures. Here we utilize the option of nucleating a skyrmion by local injection of a spin-polarized current. The write head employs the spin-valve model, in which we configured the dependence of the nucleation delay versus current density with the relationship shown in Figure 5. A larger current density can achieve faster nucleation speed, therefore there is a design trade-off between the nucleation energy and speed. The skyrmion motion along the nanotrack considered here is driven by a vertical spin current, which is injected from the heavy-metal layer with spin polarization along the $y$-direction because of the spin-orbit scattering. The skyrmion motion model can be described by solving the Thiele equation [17, 19, 28],

$$\boldsymbol{G} \times \mathbf{v}_d - \boldsymbol{\mathcal{D}}\alpha\mathbf{v}_d + \nabla V(\boldsymbol{r}) = \boldsymbol{0} \tag{1}$$

where $\mathbf{v}_d$ is the velocity of the skyrmion. The first term in the left side of equation (1) is the Magnus force with $\boldsymbol{G} = (0,0,\mathcal{G})$ as the gyromagnetic coupling vector. The second term is the dissipative force with $\boldsymbol{\mathcal{D}}$ as the dissipative force tensor. The third term $\nabla V(\boldsymbol{r})$ represents the force exerted on the moving skyrmion (*e.g.*, potential at the spatial coordinates $\boldsymbol{r}$ from the surrounding environment). The VCMA gate is implemented with a voltage-controlled switch using look-up table based on the working window shown in Figure 2b. At the read head, the presence of a skyrmion results in local tunnel conductance (or resistance) change due to the spin-mixing from the inhomogeneous magnetic non-collinearity and spin-orbit interaction [20, 29]. This conductance change can be detected with a perpendicular-to-plane current through a MTJ device. The tunnel conductance change ($\Delta G$) ratio induced by the presence of a skyrmion can be expressed as $\Delta G/G_0 \cong \pi\varepsilon^2/A$, where $G_0$ denotes the maximum tunnel conductance without any skyrmions, $A$ is the sectional area of the MTJ read head and $\varepsilon$ is the skyrmion diameter, respectively. The skyrmion diameter is proportional to the ratio between the DMI and the magnetic anisotropy, *i.e.*, $\varepsilon \sim D/K_u$. For the default parameter values, we get $\varepsilon \cong 15$ nm and the $\Delta G/G_0$ value (or TMR ratio) is about 14%. This Sky-RM model is written in the Verilog-A language [30] and is compatible with the standard CMOS design tools (*e.g.* Cadence platform [31]). Finally, the Sky-RM model is calibrated with the micromagnetic simulation results.

Based on the developed Sky-RM electrical model, we performed hybrid skyrmion/CMOS electrical simulations. Figure 6 shows the transient simulation waveforms. Under the nucleation current $I_{nucl} \cong 2.85$ mA, provided by the spin-valve write head (see Figure 6a), the binary data information (i.e., '0' or '1') is encoded by the absence or presence of a skyrmion and written into the nanotrack (see Figure 6b, e.g., with data bit sequence "…0010001100…"). Here the skyrmion nucleation delay is about 36 ps (see the inset of Figure 6). Then the skyrmion moves along the nanotrack by the driving current $I_{driv} \cong 16$ μA with pulse duration of 2 ns (see Figure 6c). The skyrmion motion velocity is about 50 m/s. As the driving current flows through the heavy metal (with low resistivity, *e.g.*, $\sim 0.2$ μΩ·m for Pt), the driving power is expected to be very small ($\sim$ nJ/m). During the motion, the skyrmion can be pinned/depinned by the VCMA gate at $x = 500$ nm (see Figure 6d). Finally the skyrmions reach the MTJ read head (see Figure 6e) and are detected by measuring the conductance change (see Figure 6f). If a skyrmion presents at the MTJ read head, the conductance will decrease, thereby the detection current will decrease, given a specific bias voltage. In our simulation, the bias voltage is set to be 0.1 V, then the detection current is $I_{det} \cong 12.57$ μA or 11.06 μA for the absence or presence of a skyrmion. Based on the current change ratio, we can also evaluate the TMR ratio. Here the evaluated TMR ratio is about 13.6%, which is a little smaller than the default value (14%) due to the bias-dependent TMR reduction [32]. As can be seen from Figure 6f, the detected data pattern at the read head is consistent with the one that sent from the write head.

**Discussions**

We have shown that a skyrmion can be pinned/depinned by using the VCMA effect. We have also presented two design strategies to employ this effect for Sky-RM application. It is worth noting that the skyrmion motion velocity is proportional to the PMA (see Supplementary Figure 2b). Therefore different design optimizations should be utilized for improving the Sky-RM performance based on the actual configuration. Specifically, if positive voltage is applied for the VCMA gate, *i.e.*, $K_{uv} > K_u$, then the width of the VCMA-gated region should be as narrow as possible to avoid hampering the skyrmion motion velocity (see Figure 7a); Alternatively, if negative voltage is applied for the VCMA gate, i.e., $K_{uv} < K_u$, then the width of the VCMA-gated region can be reasonably widen to accelerate the skyrmion motion velocity (see Figure 7b). For example, if the width of the VCMA-gated region reduces from 60 nm to 20 nm, then the time required to pass the same distance (e.g., from $x = 35$ nm to $140$ nm, which covers one VCMA gate ) changes. Specifically, the required time grows (decreases) from ~2.68 ns (~3.48 ns) to ~2.82 ns (~3.20 ns) for $K_{uv} = 0.95 K_u$ ($K_{uv} = 1.05 K_u$) with the same driving current density of $j_{driv} = 4 \text{ MA/cm}^2$, as shown in Figure 7c and Figure 7d respectively. In practice, negative-biased VCMA gate is preferable for high-speed Sky-RM applications.

As the skyrmion has a transverse velocity when it moves along the nanotrack, it may reach the nanotrack edge and be annihilated (see Supplementary Figure 2b). Therefore the driving current density should be limited in practical implementation for a steady motion along the nanotrack. Interestingly, our simulation shows that if the driving current is off, the skyrmion will move away from the edge due to the repulsive interaction on the edge. This inspires us to drive the skyrmion with a current pulse sequence instead of a DC current. As shown in Figure 8a, with the driving current density of $j_{driv} = 6 \text{ MA/cm}^2$, if the pulse width is smaller than 2 ns, the skyrmion moves successfully along the nanotrack (see Supplementary Movie 5). Otherwise, the skyrmion reaches the nanotrack edge and is eventually annihilated (e.g., with pulse width of 3 ns, see Figure 8b and Supplementary Movie 6). This provides design freedoms to modulate the skyrmion motion trajectory, speed and power by controlling the driving current pulse width. In addition, the storage density mainly depends on the nanotrack width. Therefore, there is a trade-off between the data access speed (i.e., skyrmion motion velocity) and the storage density. This trade-off is very important for the Sky-RM design and optimization. For example, we have to sacrifice the storage density (increasing the nanotrack width) to get high skyrmion motion velocity (using large driving current density) for applications (e.g., computing) requiring high-speed data access. Alternatively, we can greatly reduce the driving current density for dense packing density in mass storage applications. Figure 8c shows the design trade-off between the skyrmion longitudinal velocity and the minimum required nanotrack width with respect to the driving current density.

Another interesting observation is that the diameter of the skyrmion also depends on the PMA, *i.e.*, $\varepsilon \sim D/K_u$. Based on this observation, we can optimize the design of the read head for improving the detection sensitivity. The skyrmion detection utilizes the conductance change at the read head, *i.e.*, $\Delta G/G_0 \cong \pi \varepsilon^2 / A$, as discussed in the "Results" section. A larger skyrmion diameter can achieve better detection sensitivity. In this case, we suggest to utilize negative voltage on the MTJ read head to detect the skyrmion, as a negative voltage results in decrease of $K_u$. In addition, the dimension configuration of the read head is also critical for the skyrmion detection sensitivity. Assume that the read head is located in the central path of the nanotrack. On one hand, as the skyrmion size is relatively small (~15-20 nm) compared to the width of the nanotrack, the dimension of the read head should be large enough to capture the skyrmion (*i.e.*, cover most of the skyrmion area). On the other hand, since $\Delta G/G_0 \cong \pi\varepsilon^2/A$, the dimension (or area $A$) of the read head cannot be too large for high detection sensitivity. Therefore there is an optimum dimension of the read head for a given nanotrack width and skyrmion size. For the Co/Pt films with the nanotrack width of 80 nm, we found that the optimum diameter of the read head is about 40 nm.

In conclusion, we have investigated the voltage controlled skyrmion motion and its application for RM. We have illustrated the pinning/depinning of skyrmion by using the VCMA effect under the well-established micromagnetic frameworks. Some interesting observations and guidelines have been identified to optimize the Sky-RM design. We have developed a compact Sky-RM electrical model and performed hybrid skyrmion/CMOS evaluations with the standard electronic design automation (EDA) platform. Furthermore, design strategies and optimizations have also been proposed to promote the development of Sky-RM.

**Methods**

**Micromagnetic simulations**. The micromagnetic simulations were performed by using the well-established public domain software package Object Oriented MicroMagnetic Framework (OOMMF) with extended module of Dzyaloshinskii-Moriya interaction (DMI) [33]. The three-dimensional time-dependent magnetization dynamics is controlled by the Landau-Lifshitz-Gilbert (LLG) equation including the in-plane and out-of-plane spin transfer torques of the vertical spin current. The skyrmion nucleation process was studied in a 0.4-nm-thick cobalt platelet with an 80-nm-width square shape. The motion of skyrmion is investigated in a 0.4-nm-thick cobalt nanotrack with width of 80 nm and length of 1000 nm. The width of the voltage-controlled magnetic anisotropy (VCMA) is 60 nm. The VCMA effect is based on a linear relationship [25], *i.e.*, $K_{uv} = K_u + \vartheta V_b$, where $V_b$ is the applied voltage and $\vartheta$ is a constant coefficient. The current-induced Oersted field was also considered in the simulation. The energy density of DMI is shown as [17],

$$\varepsilon_{DM} = D(\boldsymbol{m}_z \frac{\partial \boldsymbol{m}_x}{\partial x} - \boldsymbol{m}_x \frac{\partial \boldsymbol{m}_z}{\partial x} + \boldsymbol{m}_z \frac{\partial \boldsymbol{m}_y}{\partial y} - \boldsymbol{m}_y \frac{\partial \boldsymbol{m}_z}{\partial y}) \qquad (2)$$

where $D = |\boldsymbol{D}_{12}|\sqrt{3}/al$, $\boldsymbol{D}_{12}$ is the DMI vector, $a$ is the atomic lattice constant, $l$ is the film thickness. $\boldsymbol{m}_x$, $\boldsymbol{m}_y$, $\boldsymbol{m}_z$ express the *x*, *y*, *z* components of the normalized magnetization. In our simulation, we set $D$ from 2 to 6 mJ/m².

The in-plane and out-of-plane torques induced by the vertical spin current are written as,

$$\boldsymbol{\tau}_{\text{in-plane}} = -\frac{u}{l}\boldsymbol{m} \times (\boldsymbol{m} \times \boldsymbol{m}_{\text{p}}) \qquad (3)$$

$$\boldsymbol{\tau}_{\text{out-of-plane}} = -\xi\frac{u}{l}(\boldsymbol{m} \times \boldsymbol{m}_{\text{p}}) \qquad (4)$$

where $u = \gamma_0(\hbar jP/2\mu_0 eM_s)$, $\gamma_0$ is the gyromagnetic ratio, $j$ is the current density, $P$ is the spin polarization, $e$ is the charge of electron, $\mu_0$ is the vacuum permeability, $M_s$ is the saturation magnetization, $\xi$ is the amplitude of the out-of-plane torque relative to the in-plane one, $\boldsymbol{m} = \boldsymbol{M}/M_s$ is the reduced magnetization and $\boldsymbol{m}_\mathrm{p}$ is the current polarization vector. It should be mentioned that we have set $\xi = 0$ in all simulations to remove the field-like out-of-plane torque for simplicity. Thus, the LLG equation augmented with the in-plane and out-of-plane spin current torques employed in our simulations reads:

$$\frac{d\boldsymbol{m}}{dt} = -\gamma_0 \boldsymbol{m}\times \boldsymbol{h}_\mathrm{eff} + \alpha\left(\boldsymbol{m}\times\frac{d\boldsymbol{m}}{dt}\right) - \frac{\gamma_0 \hbar jP}{2\mu_0 eM_s l}[\boldsymbol{m}\times(\boldsymbol{m}\times\boldsymbol{m}_\mathrm{p})] - \frac{\xi\gamma_0\hbar jP}{2\mu_0 eM_s l}(\boldsymbol{m}\times\boldsymbol{m}_\mathrm{p}) \quad (5)$$

where $\boldsymbol{h}_\mathrm{eff}$ is the reduced effective field.

We adopt the magnetic material parameters from Ref. [17]: the exchange stiffness $A = 15\ \mathrm{pJ/m}$, the saturation magnetization $M_s = 580\ \mathrm{kA/m}$, perpendicular magnetic anisotropy constant $K_u = 0.8\ \mathrm{MJ/m^3}$, gyromagnetic ratio $\gamma_0 = 2.21\times 10^5$, current spin polarization rate $P = 0.4$. In the simulation of the skyrmion nucleation, we consider the Gilbert damping as $\alpha = 0.3$ for Co/Pt system [27]. During the simulation of the current-driven skyrmion motion, the spin Hall angle ($\eta$) is set to be 0.25, and the PMA constant $K_u$ varies from 0.7 $\mathrm{MJ/m^3}$ to 0.9 $\mathrm{MJ/m^3}$. The PMA of the VCMA gate $K_{uv}$ varies from $0.9K_u$ to $1.1K_u$. The cell size of the discretization mesh is set to be 1 nm × 1 nm × 0.4 nm. All the parameters are in SI units.

**Electrical simulations**. The hybrid skyrmion/CMOS electrical simulations were performed on the standard Cadence platform with the Sky-RM model being implemented with the Verilog-A language [30]. The Sky-RM model is composed of for parts: spin-valve write head, nanotrack, VCMA gate and MTJ read head. The spin-valve write head is with diameter of 20 nm and spin polarization factor of $P = 0.4$. The relationship between the nucleation delay ($t_{nucl}$) and current density ($j_{nucl}$) is featured using the fitting function as,

$$t_{nucl} = t_0 + \alpha\cdot\exp(-\Delta(j_{nucl} - J_0)) \quad (6)$$

where $t_0$, $\alpha$ and $\Delta$ are fitting factors, $J_0$ is the critical current density. The nanotrack uses the Co/Pt thin films with some critical parameters as following: Gilbert damping $\alpha = 0.3$, DMI $D = 3\ \mathrm{mJ/m^2}$, perpendicular magnetic anisotropy $K = 0.8\ \mathrm{MJ/m^3}$, and size (length × width × thickness) of (1000 nm × 80 nm × 0.4 nm). The skyrmion motion (along the nanotrack) model is described by solving the Thiele equation [19]. The width of the VCMA gate is 60 nm and the PMA is expressed as, $K_{uv} = K_u + \vartheta V_b$, where the default $K_u = 0.8\ \mathrm{MJ/m^3}$, $\vartheta = 0.02\ \mathrm{MJ/V}$ and $V_b$ varies from -2 V to 2 V. The MTJ read head is with diameter of 40 nm and TMR ratio of 14%. The tunnel conductance change ($\Delta G$) ratio induced by the presence of a skyrmion can be expressed as (7) with the maximum resistance ($G_0^{-1}$) of the read head being expressed as (8),

$$\Delta G/G_0 \cong \pi\varepsilon^2/A \quad (7)$$

$$G_0^{-1} = \frac{t_{ox}}{F\times\bar{\varphi}^{1/2}\times A}\exp(1.025\times t_{ox}\times\bar{\varphi}^{1/2}) \quad (8)$$

where $F$ is a fitting factor, $\bar{\varphi} = 0.4$ is the potential barrier height of MgO barrier, $t_{ox}$ and *A* are the thickness of the MgO barrier and the sectional-area of the MTJ read head, respectively. All the peripheral circuits for generating the nucleation, driving and detection currents are formed with CMOS transistors. The CMOS transistor models are from the STMicroelectronics at 40 nm technology node [31]. The skyrmion motion velocity in the nanotrack is derived from the Thiele equation. The Sky-RM model is calibrated with the micromagnetic simulation results.

**Acknowledgements**

The authors would like to thank the supports by the projects from the Chinese Postdoctoral Science Foundation (No. 2015M570024), Chinese National Natural Science Foundation (No. 61501013, 61471015 and 61571023), French National Agency (No. ANR-MARS) and the International Collaboration Project (No. 2015DFE12880) from the Ministry of Science and Technology. Y.Z. thanks the support by the UGC Grant AoE/P-04/08 of Hong Kong SAR government. The authors also thank the fruitful discussions with Professor Albert Fert.

**Author contributions**

W.S.Z. and Y.Z. conceived and coordinated the project. Y.Q.H. carried out the micromagnetic simulations supervised by Y.Z. W.K. and C.T.Z. developed the electrical model and performed the electrical simulations. W.K., Y.Q.H., C.T.Z., W.F.L., N. L., Y.G.Z., X.C.Z., Y.Z., and W.S.Z. wrote the manuscript. All authors interpreted the data and contributed to the discussion and comment on the manuscript. Correspondence and requests for materials should be addressed to W.S.Z or Y.Z.

**Competing financial interests**

The authors declare no competing financial interests.

**Figure legends**

Figure 1. Schematic of the skyrmion-based racetrack memory (Sky-RM); (a) Front-view; (b) Top-view. It consists of five parts: spin-valve write head, nanotrack, magnetic tunnel junction (MTJ) read head, voltage-controlled magnetic anisotropy (VCMA) gate as well as peripheral complementary metal-oxide semiconductor (CMOS) circuits (which are not shown in the figure). The skyrmion is initially nucleated by injecting a local spin-polarized current ($I_{nucl}$) through the spin-valve write head (diameter $d = 20$ nm) depending on the input data, and then moves along the nanotrack (with size $1000$ nm $\times\ 80$ nm $\times\ 0.4$ nm) by the vertical spin current ($I_{driv}$). During the motion, the skyrmion can be pinned/depinned at the VCMA gate (with width of 60 nm) by modulating the magnetic anisotropy. Finally the skyrmion can be detected by applying a detection current ($I_{det}$) across the MTJ read head due to the TMR change.

Figure 2. (a) The schematic view (x-y plane) of the skymion states when passing the VCMA-gated region, where the yellow-line shadow represents the VCMA-gated region. Initial state: both the driving current and the voltage of the VCMA gate are off, the skyrmion locates at its original position. When the driving current and the VCMA gate are turned on, the skyrmion moves from the left side to the right side of the nanotrack. Three cases may occur: (i) the energy barrier is $\{+\Delta E_b, -\Delta E_b\}$ and the driving current cannot overcome $+\Delta E_b$, the skyrmion stops at the left side of the VCMA-gated region; (ii) the energy barrier is $\{-\Delta E_b, +\Delta E_b\}$ but the driving current cannot overcome $+\Delta E_b$, the skyrmion stops at the right side of the VCMA-gated region; (iii) the driving current is able to overcome $+\Delta E_b$, the skrymion passes the VCMA-gated region. (b) Illustration of the energy barriers of the VCMA-gated region. (c) Working window of the skyrmion pining/depinning states for the baseline $K_u = 0.8$ MJ/m$^3$ at various voltages (or $+\Delta E_b$) and driving current densities. (d) Working window of the skyrmion pining/depinning states for a constant energy barrier height ($|\Delta E_b| = 0.04$ MJ/m$^3$) at various baseline $K_u$ and driving current densities. The green square denoted the "pass" state, the red circle denoted the "stop" state and the black cross denotes the "broken" state, *i.e.*, the skyrmion reaches the edge of the nanotrack and is annihilated during the motion process.

Figure 3. Trajectory of the skyrmion motion along the nanotrack for the case of controlling the on/off voltage of the VCMA-gate, in which each dot denotes the center of the skyrmion at the corresponding time. Here the length of the nanotrack is set to be 300 nm for saving simulation time. (a) Positive voltage pulses to control the VCMA gate with active duration of 5 ns and idle duration of 2 ns; The energy barrier pair of each VCMA-gated region is $\{+0.05K_u, -0.05K_u\}$; (b) Negative voltage pulses to control the VCMA gate with active duration of 5 ns and idle duration of 1 ns; The energy barrier of each VCMA-gated region is $\{-0.05K_u, +0.05K_u\}$. When the voltage of the VCMA-gate is on, the skyrmion moves along the nanotrack under the DC driving current (with amplitude of $j_{driv} = 2$ MA/cm$^2$) and stops at the left (right) side of the VCMA-gated region for the case of positive (negative) voltage.

Then when the voltage of the VCMA gate is off, the skyrmion continues to move on until the next VCMA gate is active.

Figure 4. Trajectory of the skyrmion motion along the nanotrack for the case of modulating the driving current configuration, in which each dot denotes the center of the skyrmion at the corresponding time. Here the length of the nanotrack is set to be 300 nm for saving simulation time. In this case, the voltage of the VCMA gate is always active. (a) Positive constant voltage ($V_b = +2.0\,V$) to control the VCMA gate. The energy barrier of each VCMA-gated region is $\{+0.05K_u, -0.05K_u\}$ and the active AC current pulse width is 5 ns; (b) Negative constant voltage ($V_b = -2.0\,V$) to control the VCMA gate. The energy barrier of the VCMA-gated region is $\{-0.05K_u, +0.05K_u\}$ and the active AC current pulse width is 1 ns. The driving current is composed of a DC component (with amplitude of $2\,MA/cm^2$) and an AC pulsed current (with amplitude of $2\,MA/cm^2$). The skyrmion moves along the nanotrack under only the DC current but stops either at the left side (for $V_b = +2.0\,V$) or at the right side (for $V_b = -2.0\,V$) of the VCMA-gated region if the AC pulsed current is off. When the AC current is on, the total amplitude of the driving current ($4\,MA/cm^2$) is large enough to overcome the energy barrier. In this case, the skyrmion passes the VCMA-gated region successfully if the pulse duration of the AC current is long enough.

Figure 5. The dependence of the nucleation delay versus the current density. A larger current density can achieve faster nucleation speed, therefore there is a design trade-off between the nucleation power and speed.

Figure 6. Transient simulation waveforms of the Sky-RM with the first pining/depinning strategy. (a) The skyrmion nucleation current $I_{nucl} \cong 2.85$ mA, provided by the spin-valve write head. The nucleation delay is about 36 ps as shown in the inset figure. (b) The write head state. Here the binary data information (*i.e.*, '0' or '1') is encoded by the absence or presence of the skyrmion. (c) The driving current $I_{driv} \cong 16$ μA, with pulse width of 2 ns. The skyrmion motion velocity under this driving current is about 50 m/s. (d) The nanotrack state at the position of $x = 500$ nm, *i.e.*, one of the VCMA gates of the nanotrack. (e) The read head state. The skyrmions from the write head finally reach the read head. (f) The detection current at the MTJ read head for a given bias voltage (*e.g.*, 0.1 V). If there is no skyrmion (*i.e.*, for data bit '0') at the read head, the detection current is $I_{det} \cong 12.57$ μA. Otherwise, if a skyrmion presents at the read head, the resistance (conductance) will increase (decrease), thereby the detection current decreases to $I_{det} \cong 11.06$ μA. Accordingly, the TMR ratio can be calculated as 13.6%. As can be seen, the same data pattern can be detected as that sent from the write head.

Figure 7. Design optimizations for the Sky-RM. (a) For VCMA gate controlled with positive voltage, smaller VCMA-gated region is preferable in order not to hamper the skyrmion

motion velocity; (b) For VCMA gate controlled with negative voltage, wider VCMA-gated region is preferable to accelerate the skyrmion motion velocity; (c)-(d) Simulation illustrations of the above two design schemes. Here the length of the nanotrack is set to be 180 nm and each dot denotes the center of the skyrmion at the corresponding time.

Figure 8. (a) Trajectory of a skyrmion driven by a current pulse sequence with density of $j_{driv} = 6 \text{ MA/cm}^2$ and pulse width of $t_{driv} = 2$ ns; (b) with density of $j_{driv} = 6 \text{ MA/cm}^2$ and pulse width of $t_{driv} = 3$ ns; (c) The trade-off between the storage density (*i.e.*, minimum nanotrack width) and the skyrmion motion velocity with respect to the driving current density.

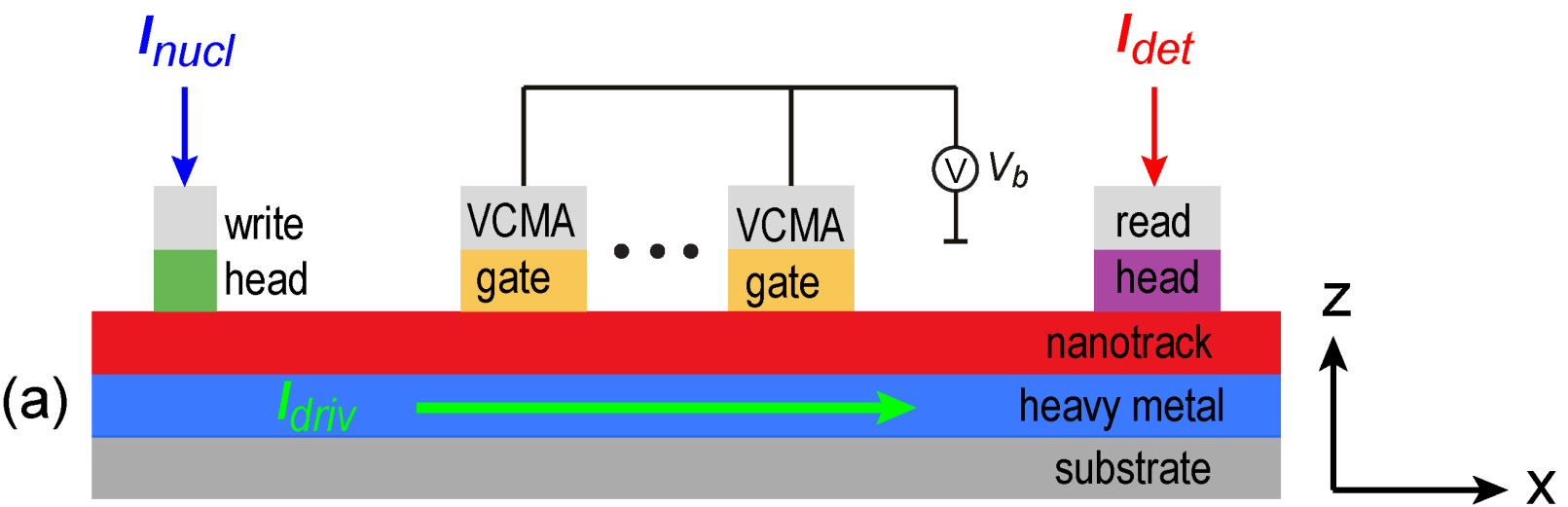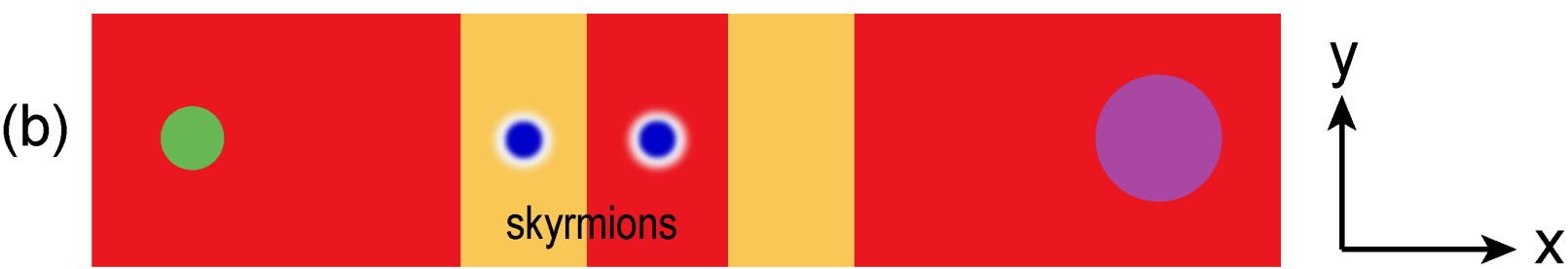

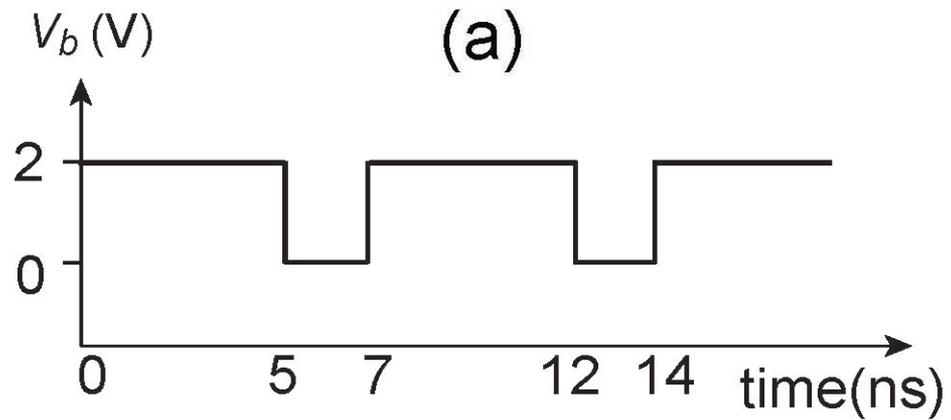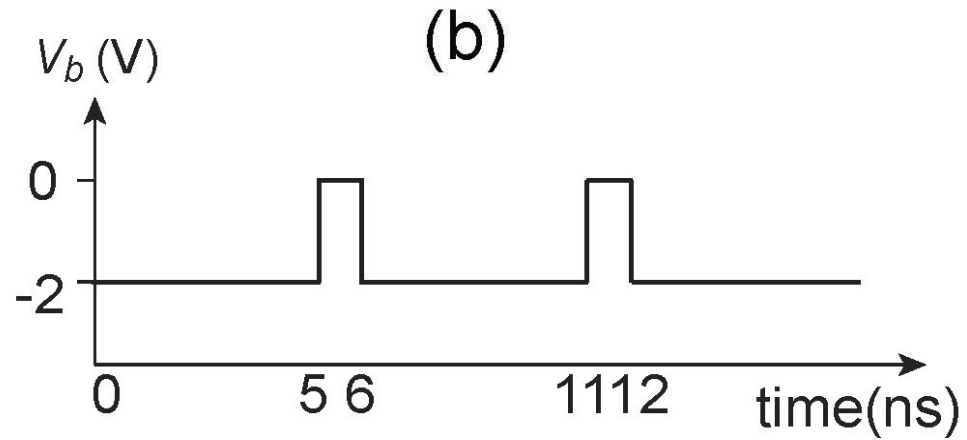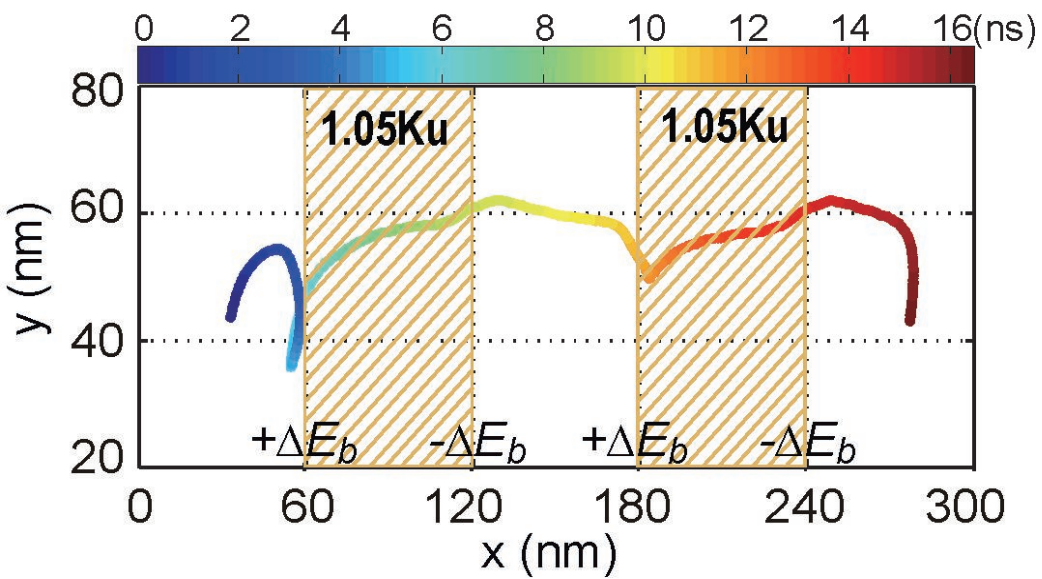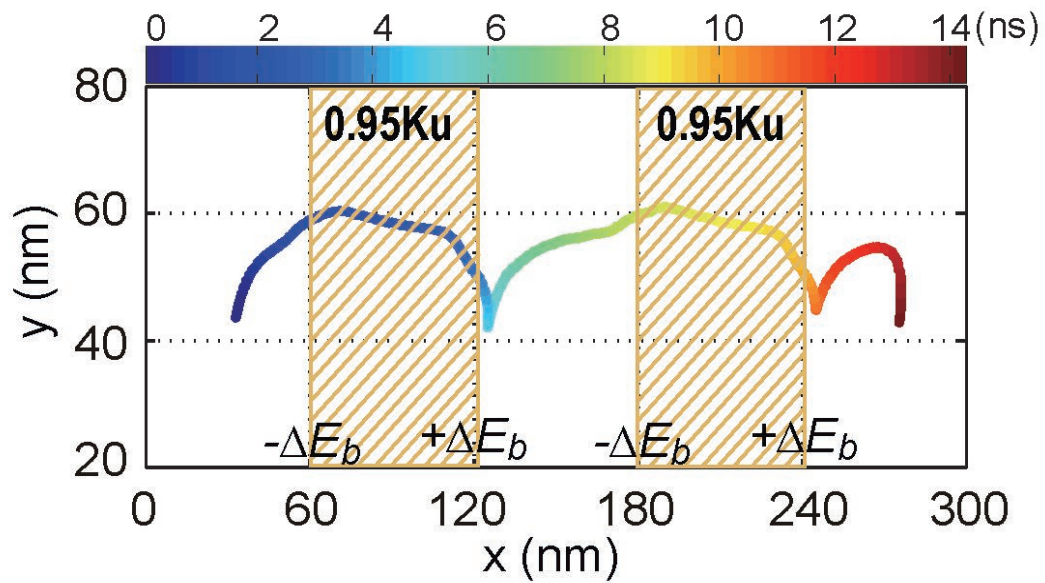

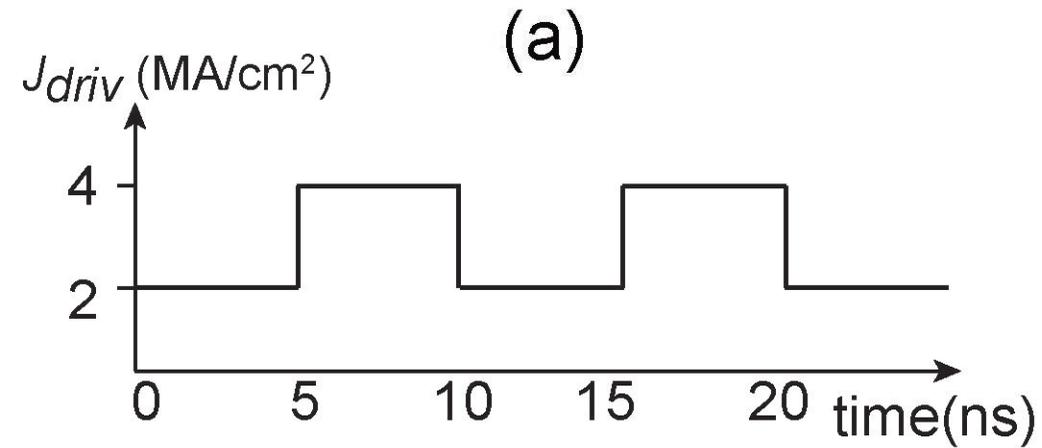
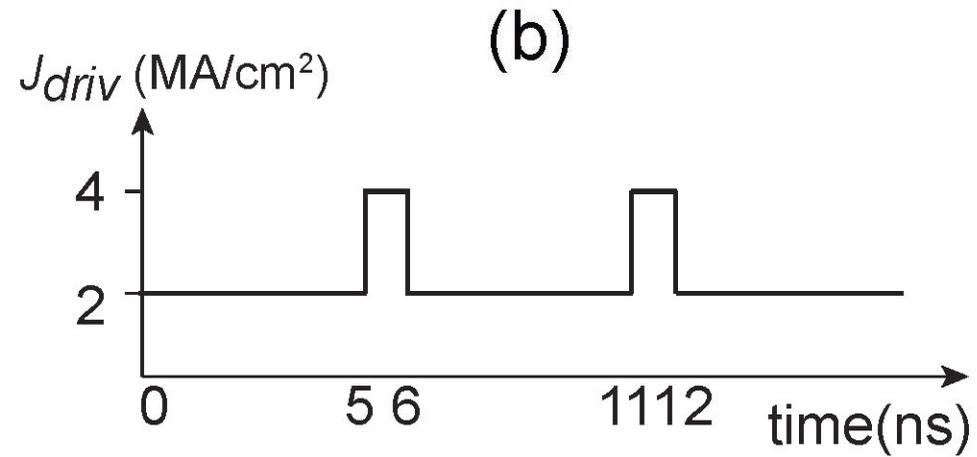

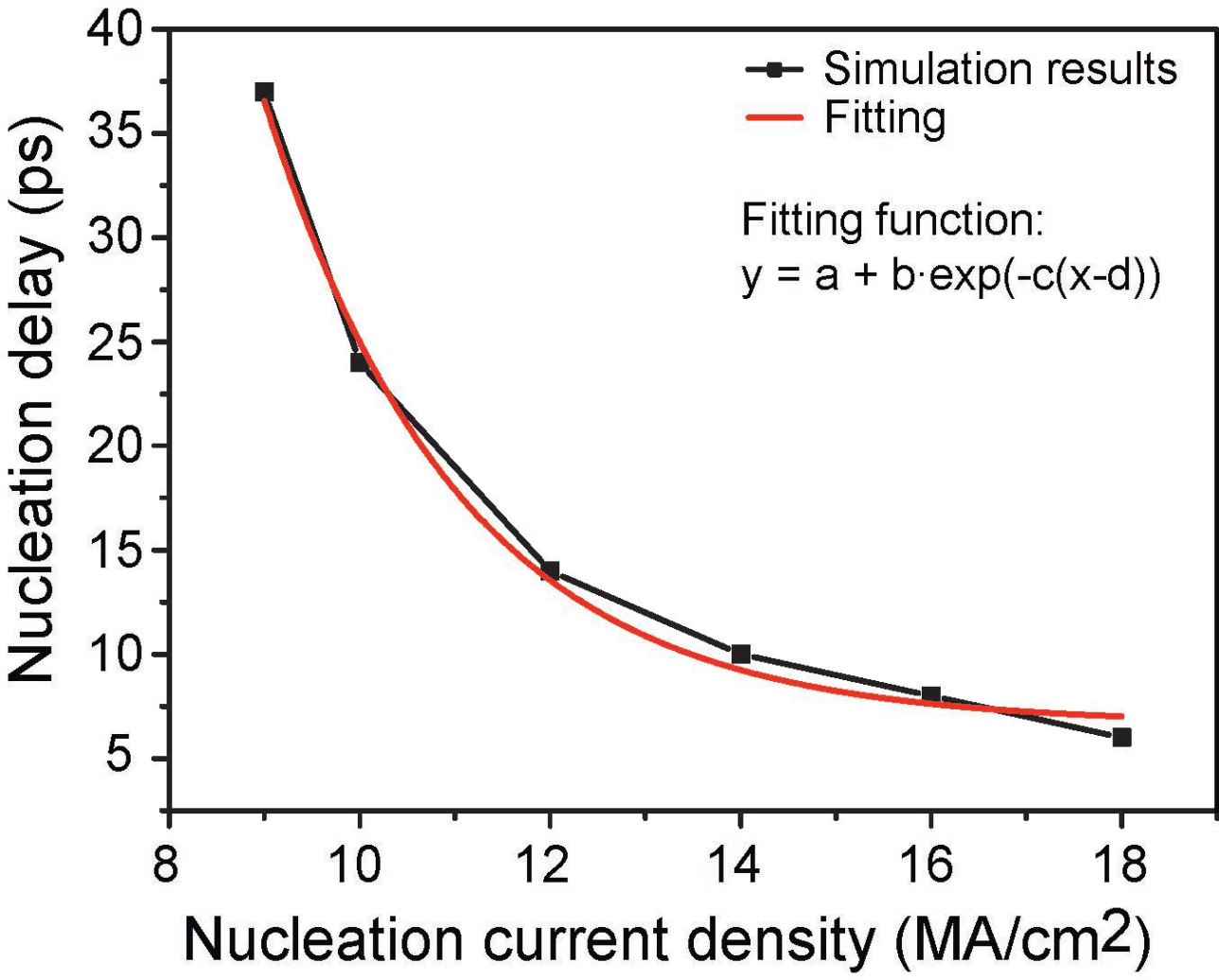

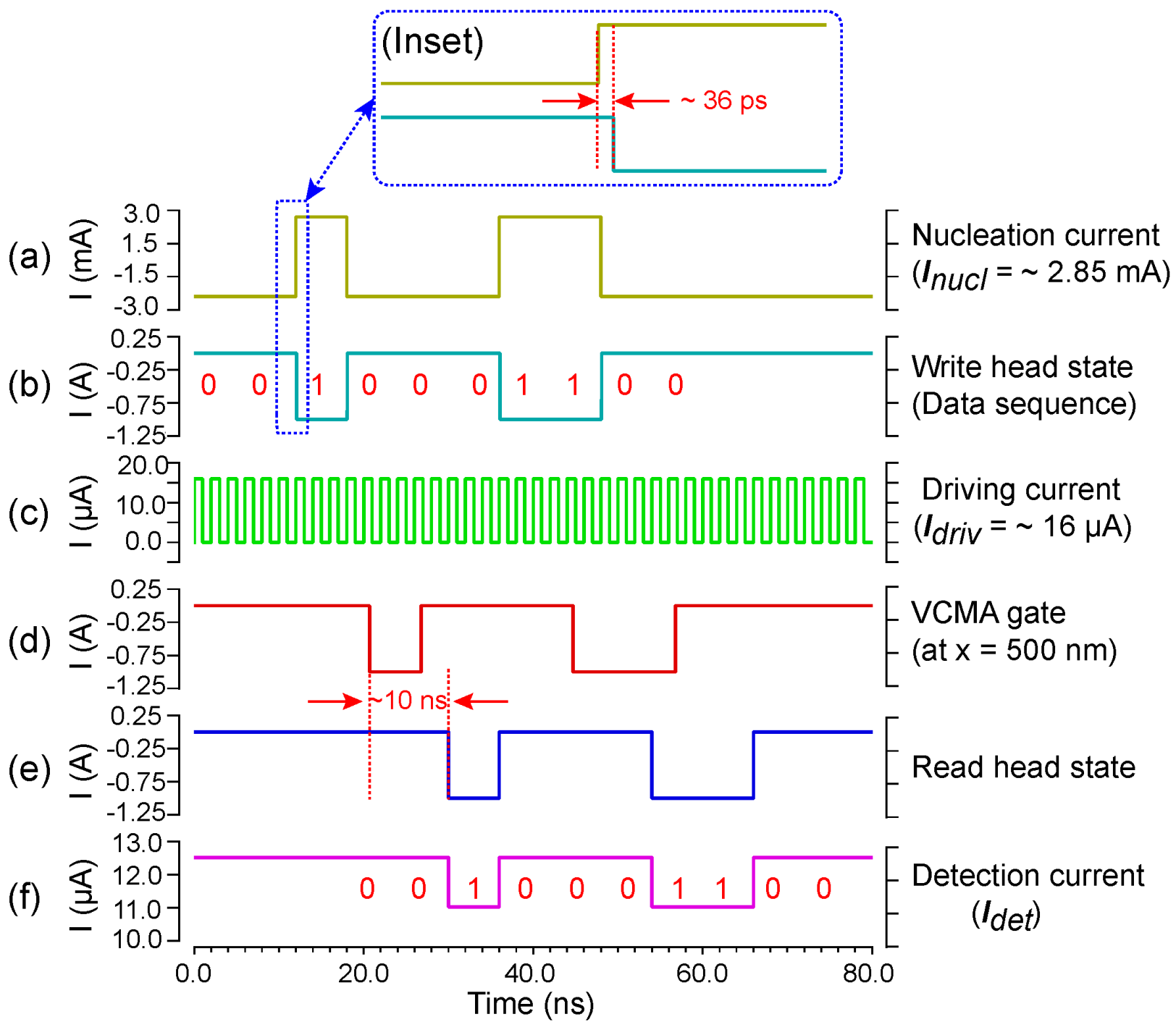

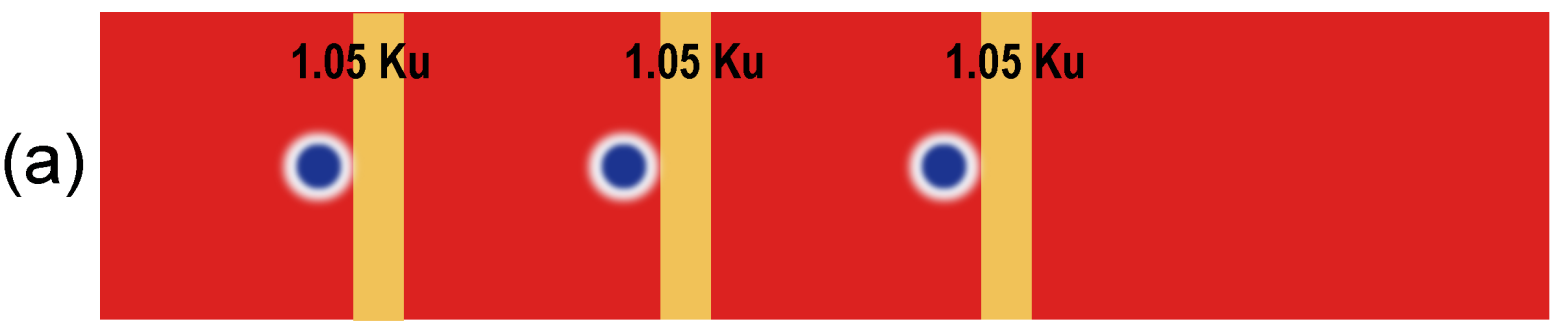
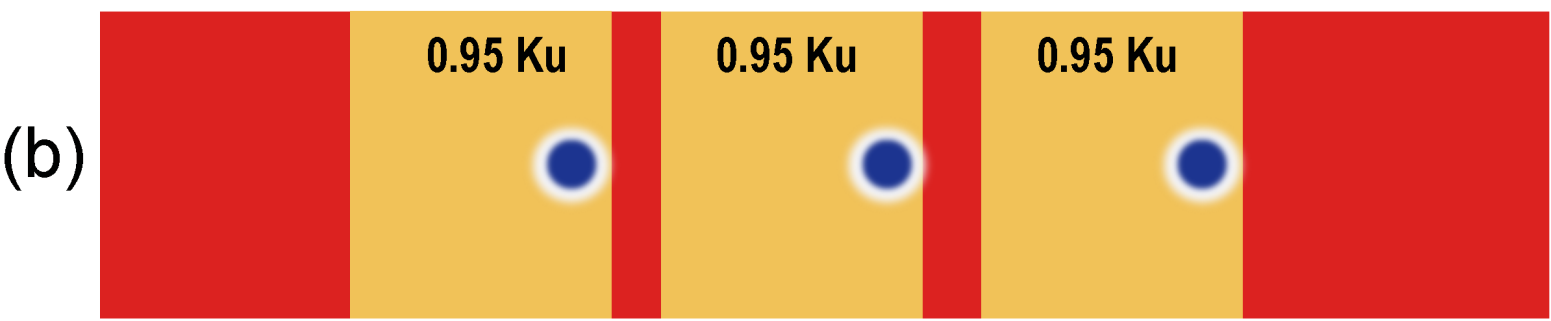
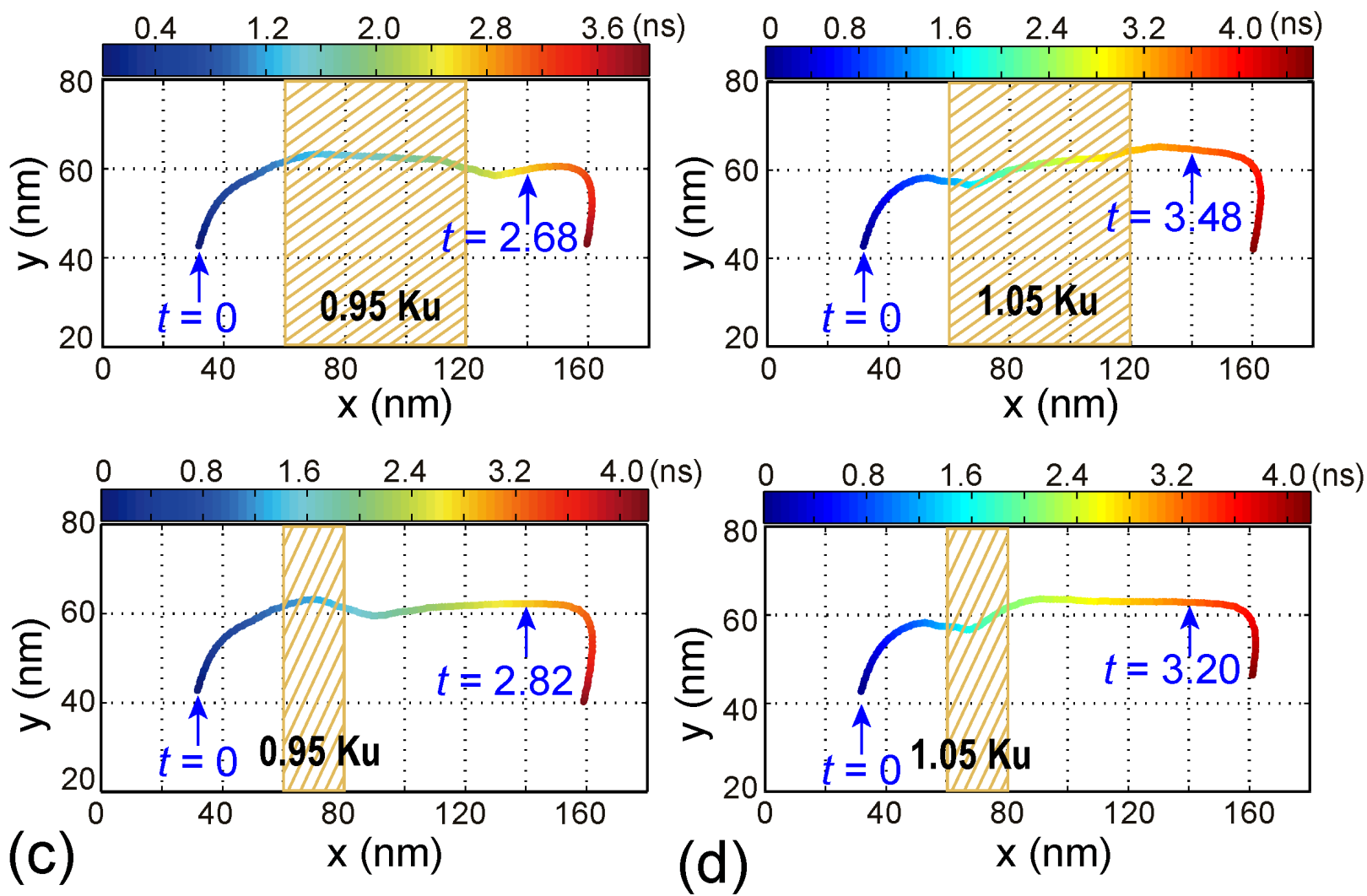

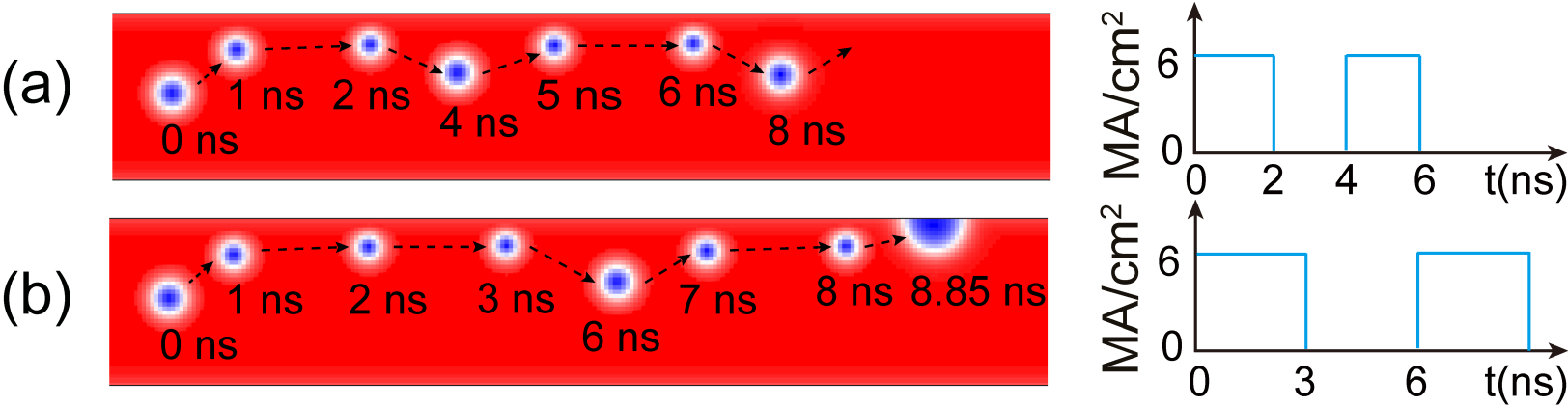
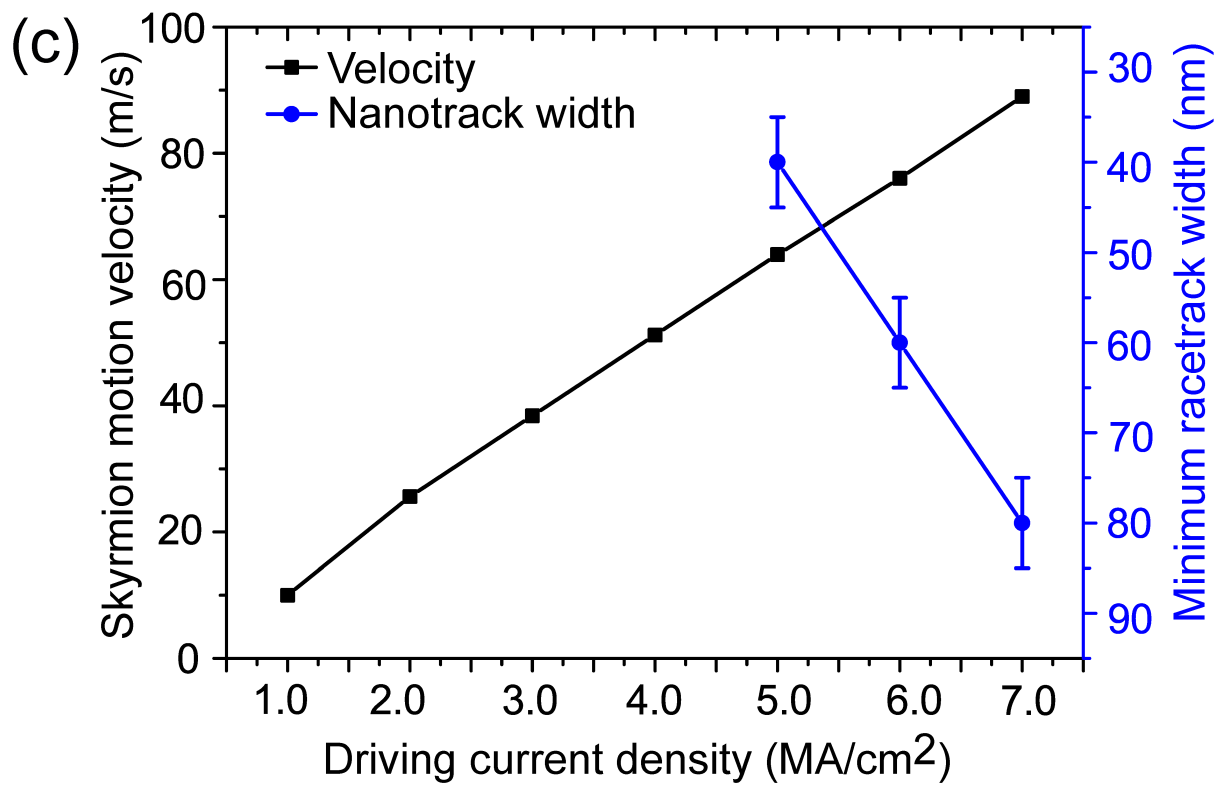